\documentclass[twocolumn,tighten]{aastex631}
\usepackage{isotope,mathrsfs}
\usepackage{amsmath,bbold,amssymb,epsfig,bm,mathrsfs,feynmp}
\usepackage{color,slashed,nicefrac,exscale,multirow,soul}
\usepackage{times,txfonts,xcolor,graphicx,gensymb,tikz}
\usepackage[normalem]{ulem}
\definecolor{red}{rgb}{0.8,0,0}
\definecolor{violet}{rgb}{0.4,0,0.4}
\definecolor{green}{rgb}{0,0.5,0.0}
\definecolor{navy}{rgb}{0.0,0.0,0.6}
\definecolor{orange}{rgb}{0.8,0.2,0.0}
\newcommand{\bea}{\begin{eqnarray}}
\newcommand{\eea}{\end{eqnarray}}
\newcommand{\ep}{\varepsilon}
\newcommand{\MR}{$M$-$R\ $}
\newcommand{\MeV}{\text{MeV}}
\newcommand{\Lsym}{\ensuremath{L_{\text{sym}}}}
\newcommand{\Qsat}{\ensuremath{Q_{\text{sat}}}}
\newcommand{\rhotran}{\ensuremath{\rho_{\text{tran}}}}

\begin{document}
\title{Relativistic hybrid stars with sequential first-order phase transitions \\
in light of multimessenger constraints}
\author[0000-0001-8635-3939]{Jia Jie Li}
\affiliation{School of Physical Science and Technology, 
             Southwest University, Chongqing 400715, China}

\author[0000-0001-9626-2643]{Armen Sedrakian}
\affiliation{Frankfurt Institute for Advanced Studies,
             D-60438 Frankfurt am Main, Germany}         
\affiliation{Institute of Theoretical Physics,
            University of Wroclaw, 50-204 Wroclaw, Poland}
\author[0000-0001-9675-7005]{Mark Alford}
\affiliation{Department of Physics, Washington University, 
             St.~Louis, MO 63130, USA}
\email{jiajieli@swu.edu.cn\\ sedrakian@fias.uni-frankfure.de\\ 
alford@physics.wustl.edu}            
\shortauthors{Li, Sedrakian \& Alford}
\begin{abstract}
In this work, we consider the properties of compact stars 
in which quark matter has low- and high-density phases that 
are separated by a first-order phase transition. Thus, unlike 
the commonly considered case of a single phase transition 
from hadronic to quark matter, our models of hybrid stars 
contain sequential phase transitions from hadronic matter 
to low- and then to high-density quark matter phases. 
We extend our previous study of the parameter space of 
hybrid stars with a single phase transition to those with 
sequential phase transitions, taking into account the 
constraints on the mass and radius of neutron stars from 
the NICER experiment, the experimental inferences of the 
neutron skin thickness of the lead nucleus by the PREX-II 
experiment, and constraints on the tidal deformability from 
the gravitational-wave event GW170817. We determine the 
range of the masses for which both twin and triplet 
configurations, i.e., identical-mass stars with two and 
three different values of radii, arise.
\end{abstract}
%
\section{Introduction}
\label{sec:Intro}
The possibility of phase transition from hadronic to quark matter 
and its implications for the properties of compact stars (CSs) has 
been the focus of researchers since early work on the subject 
several decades ago~\citep{Ivanenko1965,Itoh1970,Collins1975}; for 
recent reviews, see~\citet{Alford2008,Anglani2014,Pisarski2019}. 
An interesting special case is the possibility of a first-order 
phase transition between the hadronic and quark phases, which occurs 
when mixed phases are disfavored by surface tension and electrostatic 
energy costs~\citep{Alford2001}. It has been established since that 
the first-order phase transition leads to a softening of the equation 
of state (EoS) and lower maximum masses of the sequences of hybrid 
stars compared to their purely hadronic counterparts. In a previous 
paper~\citep{Lijj2021} we studied the implications of the recent 
radius determination of PSR J0740+6620 by the NICER experiment and 
the determination of the neutron skin by the PREX-II experiment 
(which gives information about symmetry energy and its slope) for 
the structure of hybrid stars with a strong first-order phase transition 
from nucleonic to quark matter. It was argued that if the interpretation 
of the PREX-II experiment implies a stiff EoS, and hence a large radius 
for the nucleonic branch, an early first-order phase transition may 
relax the tensions with the astrophysical inferences of radii for the 
relevant ranges of masses of CSs. We also deduced the ranges of mass 
and radius of twin stars which, however, were restricted to a relatively 
narrow domain of masses and radii. 

The purpose of this work is to implement the idea of sequential 
first-order phase transitions in a hybrid star~\citep{Alford2017}, 
in particular the case where there are two such transitions, 
which can be realized in Nambu--Jona-Lasinio (NJL) 
models~\citep[e.g.,][]{Blaschke2010,Bonanno2012,Blaschke2013}. 
Here, we will continue using the scheme developed in our previous 
work~\citep{Lijj2020a,Lijj2021} which accounts for the currently 
available multimessenger information coming from various channels 
of astronomical observations and terrestrial experiments. Let us give, 
at this point, a brief list of the constraints against which our models 
developed below will be tested:
\begin{itemize}
\item PSR J0030+0451:
This is the first object with mass {\it and} radius inferred to a 
precision of around 10\%~\citep{NICER2019a,NICER2019b}. These quantities 
were extracted from the fits to the data coming from the NICER observatory, 
which required modeling the soft X-ray pulses produced by the stellar 
rotation and the hot spots on the star's surface and fitting to the 
waveforms. The two (independent) analyses predict 
($68\%$ credible interval) $M=1.34^{+0.15}_{-0.16} M_{\odot}$, 
$R= 12.71^{+1.14}_{-1.19}$~km~\citep{NICER2019a}
and $M = 1.44^{+0.15}_{-0.14} M_{\odot}$, 
$R = 13.02^{+1.24}_{^-1.06}$~km~\citep{NICER2019b}. 
\item PSR J0740+6620: 
The mass of this pulsar was initially measured~\citep{NANOGrav2019} 
using Shapiro delay to be $2.08^{+0.07}_{-0.07}\,M_\odot$. Its mass 
and radius were determined using the NICER X-ray light curves with 
the results for the radius $12.39^{+1.30}_{-0.98}$~\citep{NICER2021a} 
and $13.71^{+2.61}_{-1.50}$ km~\citep{NICER2021b} and corresponding 
mass estimates $2.07^{+0.07}_{-0.07}\,M_{\odot}$ and
$2.08^{+0.09}_{-0.09}\,M_{\odot}$ ($68\%$ credible interval). 
\item GW170817: The tidal deformability (TD) of a star of mass
$\sim 1.4\,M_\odot$ in the GW170817 event by the LIGO-Virgo 
Collaboration~\citep{LIGO_Virgo2017,LIGO_Virgo2018,LIGO_Virgo2019} 
was constrained to be below the (dimensionless) value 
$\Lambda_{1.4} \le 580$, which implies a soft EoS at the 
intermediate (a few times nuclear saturation) density range. 
This observation is consistent with a phase transition from 
hadronic to quark matter, which can improve the agreement of 
theoretical models with the TD extracted from GW170817 
event~\citep[e.g.,][]{Annala2018,Paschalidis2018,Most2018,Tews2018,
Burgio2018,Alvarez-Castillo2019,Christian2019,Montana2019,
Sieniawska2019,Essick2019,Lijj2020a,Miaozq2020,Liang2021,
Malfatti2020,Rodriguez2021,Tan2022}. Furthermore, the phase 
transition may lead to the emergence of new branches of stable 
CSs, which provides a new diagnostics of phase transition in CSs, 
through observation of {\it twin and triplet stars}, i.e., 
one or two hybrid stars having the same mass, but different radii 
from, a purely hadronic 
star~\citep[e.g.,][]{Alford2013,Alford2017,Paschalidis2018,
Alvarez-Castillo2019,Christian2019,Montana2019,Lijj2020a,Christian2021}. 
\end{itemize}

Due to the high mass of PSR J0740+6620, it probes the region of 
densities that is highly relevant to a possible phase transition 
to quark matter. Recent work on the EoS has generally concluded, 
independent of the details of the EoS and methods of comparison 
with the multimessenger data adopted, that the EoS of the star 
must be moderately soft at intermediate densities and stiff enough 
at high densities. The first requirement accounts for the small TD 
in GW170817, while the second one accounts for the large mass of 
PSR J0740+6620~\citep[e.g.,][]{Lijj2021,Tanhuang2021,Biswas2021,
Legred2021,Raaijmakers2021,Huth2021,Zhangnb2021,Tangsp2021,
Christian2021,Jokela2021,Drischler2021,Contrera2022}.

The isospin dependence of the nuclear interaction is constrained 
by the measurements of the neutron skin of nuclei. Most recently, 
the Lead Radius Experiment Collaboration (PREX-II) measured the 
neutron skin thickness of the lead nucleus, 
$R^{208}_{\rm skin}= 0.283 \pm 0.071$~fm 
(mean and $1\,\sigma$ deviation), in a parity-violating 
electron-scattering experiment~\citep{PREX-II2021}. The existing theoretical 
analyses~\citep{Reed2021,Reinhard2021} do not converge to a consistent 
mean value of the symmetry energy, $E_{\rm sym}$, and its slope, 
$L_{\rm sym}$, at nuclear saturation density. While the first 
reference~\citep{Reed2021} infers $E_{\rm sym} = 38.1 \pm 4.7$~MeV 
and $L_{\rm sym} = 106 \pm 37$~MeV using a family of relativistic 
density functionals (DFs), the second reference~\citep{Reinhard2021} 
finds $E_{\rm sym} = 32 \pm 1$~MeV and $L_{\rm sym} = 54 \pm 8$~MeV 
using a larger number of relativistic and nonrelativistic DFs. 
(These two analyses are still consistent with each other at better 
than $2\sigma$ accuracy.) They also included additional constraints 
from the experimental limits on the dipole polarizability of $^{208}$Pb, 
which prefer DFs predicting a small value of 
$L_{\rm sym}$~\citep{Reinhard2021}. While the second set of parameter 
values is within the standard range~\citep{Lattimer2013,Danielewicz2014,
Oertel2017,BaldoBurgio2016}, the first set is not. In particular, their 
large value of $L_{\rm sym}$ is in tension with the GW170817 deformability 
measurement if one assumes a purely nucleonic composition~\citep{Reed2021}.
Other authors~\citep{Essick2021} have confirmed the tension between 
the results of~\citet{Reed2021} and astrophysical data using 
a nonparametric EoS. Thus, following our previous work~\citep{Lijj2021}, 
we will consider a broader range of $L_{\rm sym}$ to cover the possibilities 
claimed in the work above. 

The physics of dense quark matter allows for multiple quark phases with 
distinct properties. One standard scenario is two-flavor color-superconducting 
(2SC) at medium densities and the color-flavor-locked (CFL) phase at high 
density. Thus, one of the aims of this work is to expand on the previous 
study~\citep{Lijj2021}, which was restricted to twin configurations, by 
considering triplets---three stars with the same mass but different 
radii---which can arise if there are three disconnected branches in the 
mass-radius (hereafter $M$-$R$) diagram~\citep{Alford2017,Lijj2020a}.

The paper is organized as follows. In Section~\ref{sec:Construction} 
we briefly define the EoS that we use to describe the hadronic and quark
phases. In Section~\ref{sec:Results} we assess the existence of twin or 
triplet configurations and confront the resultant EoS models with the 
inferences from GW170817 and NICER observations of PSR J0740+6620. 
Our conclusions are given in Section~\ref{sec:Conclusions}.

\section{Construction of equation of state and basic picture}
\label{sec:Construction}
%
\subsection{Nuclear matter equation of state}
\begin{figure}[tb]
\centering
\includegraphics[width = 0.45\textwidth]{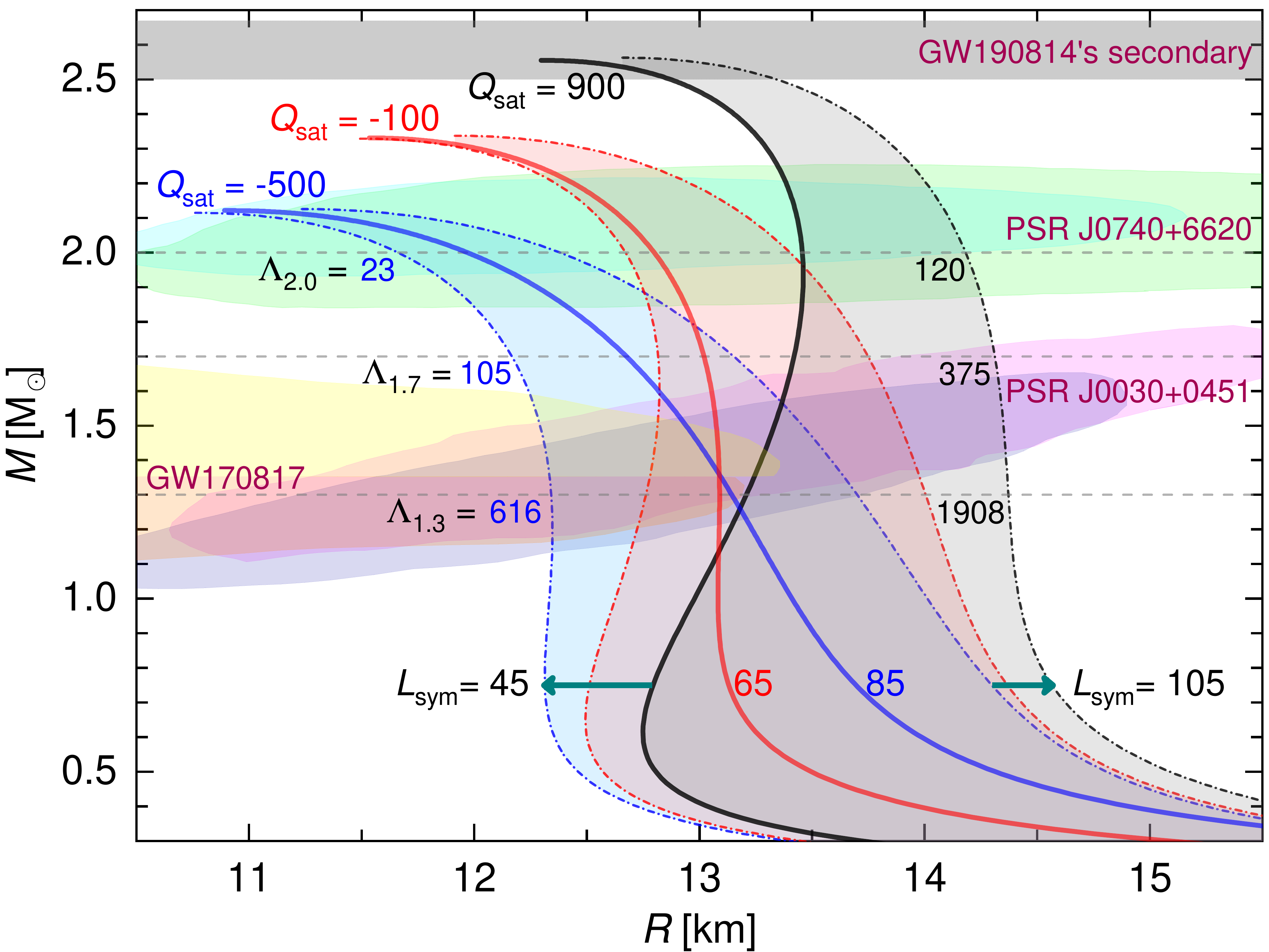}
\caption{
\MR relation for nucleonic EoS within different pairs of 
values of $Q_{\rm sat}$ and $L_{\rm sym}$ (in MeV). We show 
three ranges of \MR curves, for $Q_\text{sat}=900\,\MeV$ (gray), 
$-100\,\MeV$ (red) and $-500\,\MeV$ (blue). For each, 
$L_{\rm sym}$ is varied from $45\,\MeV$ to $105\,\MeV$. 
Thick solid lines show combinations of ($Q_{\rm sat}, L_{\rm sym}$) 
that could marginally meet the GW170817 constraint, i.e., 
(900, 45), (-100, 65), and (-500, 85). In addition, we show 
the TDs of CSs with mass $M/M_{\odot} = 1.3, 1.7$, and 2.0 for 
two extreme models. Constraints at 90\% credibility from 
multimessenger astronomy are shown by shaded 
regions~\citep{LIGO_Virgo2019,NICER2019a,NICER2021a,NICER2019b,NICER2021b}; 
see text for details.}
\label{fig:TempoN}
\end{figure}

We will adopt for the hadronic phase the same description as in the 
predecessor paper~\citep{Lijj2021}. We provide below a brief account 
of our approach for the sake of completeness. The hadronic matter is 
described within a covariant density-functional (CDF) approach with 
density-dependent nucleon-meson couplings~\citep{Lalazissis2005}. 
The density dependence of the coupling allows us to establish a one-to-one 
correspondence between our CDF and the purely phenomenological expansion 
of the energy density of nuclear 
matter~\citep[e.g.,][]{Margueron2018} in the vicinity of the saturation 
density, $\rho_{\rm sat}$, with respect to the number density $\rho$ and 
isospin asymmetry $\delta = (\rho_{\rm n}-\rho_{\rm p})/\rho$ where 
$\rho_{\rm n(p)}$ is the neutron (proton) number density, as
\bea
\label{eq:Taylor_expansion}
E(\chi, \delta) & \simeq & E_{\rm{sat}} + \frac{1}{2!}K_{\rm{sat}}\chi^2
                           + \frac{1}{3!}Q_{\rm{sat}}\chi^3 \nonumber \\ [1.0ex]
                &  & +\,E_{\rm{sym}}\delta^2 + L_{\rm{sym}}\delta^2\chi
                           + {\mathcal O}(\chi^4, \chi^2\delta^2),
\eea
where $\chi=(\rho-\rho_{\rm{sat}})/3\rho_{\rm{sat}}$.
The coefficients of this double expansion are referred to commonly 
as {\it incompressibility}, $K_{\rm sat}$, the {\it skewness}, 
$Q_{\rm{sat}}$, the {\it symmetry energy}, $E_{\rm{sym}}$, and 
its {\it slope parameter}, $L_{\rm{sym}}$. The mapping between 
the CDF and the phenomenological expansion~\eqref{eq:Taylor_expansion}
allows us to express the gross properties of CSs in terms of 
physically transparent quantities. 

In this work, we use three sets of representative EoS models,
taking three values of $Q_{\rm sat} = -500, -100$, and 900\,MeV, and
exploring values of $L_{\rm sym}$ ranging from 45 to $105$\,MeV. 
Larger values of $L_{\rm sym}$ correspond to a stiffer EoS near 
nuclear saturation density, leading to larger radii for a $1.4\,M_\odot$ 
star. Larger values of $Q_{\rm sat}$ mean that the EoS is stiffer 
at high density, thereby increasing the maximum mass of a static 
nucleonic CS~\citep[e.g.,][]{Lijj2019b,Zhangnb2018,Margueron2018}.
For $Q_{\rm sat} = -500$\,MeV the maximum mass is about $2.1\,M_\odot$, 
which matches the mass measurement of 
PSR J0740+6620~\citep{NANOGrav2019,Fonseca2021}; 
for $Q_{\rm sat} = -100$\,MeV the maximum mass is consistent with 
the (approximate) {\it upper limit} on the maximum mass of static 
CSs $\sim2.3\,M_\odot$ inferred from the analysis of the GW170817 
event~\citep{Rezzolla2018,Khadkikar2021}; finally, for 
$Q_{\rm sat} = 900$\,MeV the maximum mass is close to
$2.5\,M_\odot$, which would be compatible with the mass of the 
secondary in the GW190814 event~\citep{LIGO-Virgo2020} and its 
interpretation as a nucleonic CS~\citep{Fattoyev2020,Sedrakian2020,Lijj2020b}. 
We choose the range of $L_{\rm sym}$ between the central value and 
the lower limit of the 90\% credible interval (CI) of the PREX-II 
measurement~\citep{PREX-II2021,Reed2021}. 

The \MR relations for our nucleonic EoS models are shown in 
Figure~\ref{fig:TempoN}, along with the current astrophysical 
observational constraints. These include
(i) the ellipses obtained by the two NICER modeling groups for PSR~J0030+0451 
and J0740+6620~\citep{NICER2019a,NICER2021a,NICER2019b,NICER2021b}; 
(ii) the regions for each of the two CSs that merged in the 
gravitational-wave (GW) event GW170817~\citep{LIGO_Virgo2019}; and 
(iii) the mass of the secondary component of GW190814~\citep{LIGO-Virgo2020}. 
All the regions/limits are given at 90\% CI. As seen from 
Figure~\ref{fig:TempoN}, the softness of the EoS (as implied by the 
GW170817 event) and stiffness at low and intermediate densities implied 
by the large value of $L_{\rm sym}$ suggested by one of the 
analyses~\citep{Reed2021} of the PREX-II experiment can be reconciled by 
an appropriate choice of the parameters. Indeed, this could be accomplished 
by nucleonic EoS that trade stiffness at high density for softness at 
low density, e.g., $L_{\rm sym} \lesssim 85$\,MeV
for $Q_{\rm sat} \sim -500$\,MeV and $L_{\rm sym} \lesssim 45$\,MeV  
(which is at the lower end of the 90\% CI of~\citet{Reed2021}) 
for $Q_{\rm sat} \sim 900$\,MeV. Thus, we conclude that the nucleonic 
models are not inconsistent with the current information available from 
multimessenger astrophysics if fairly low values of $L_{\rm sym}$ are 
adopted. We next explore how this situation changes when sequential 
first-order phase transitions are allowed. 

\subsection{Quark matter equation of state}
We will model below the EoS of the quark phase using a 
{\it synthetic} constant-sound-speed (CSS) 
parameterization~\citep{Zdunik2013,Alford2013}, 
which matches well with the predictions based on the 
NJL model computations that include vector 
repulsion~\citep[e.g.,][]{Blaschke2010,Bonanno2012,Blaschke2013}.
The extension of the CSS EoS to the case of two 
sequential phase transitions is given by~\citep{Alford2017},
\bea\label{eq:EoS}
p(\ep) =\left\{
\begin{array}{ll}
p_1, &\quad  \ep_1 < \ep < \ep_1\!+\!\Delta\ep_1 \\[0.5ex]
p_1 + s_1 \bigl[\ep-(\ep_1\!+\!\Delta\ep_1)\bigr],
     &\quad \ep_1\!+\!\Delta\ep_1 < \ep < \ep_2 \\[0.5ex]
p_2, &\quad \ep_2 <\ep < \ep_2\!+\!\Delta\ep_2 \\[0.5ex]
p_2+ s_2\bigl[\ep-(\ep_2\!+\!\Delta\ep_2)\bigr], 
     &\quad \ep > \ep_2\!+\!\Delta\ep_2 \,
\end{array}
\right.
\eea
where $p_{1,2}$ and $\ep_{1,2}$ are the pressure and energy density 
at which the transition from hadronic to quark matter and from 
low-density quark phase (hereafter Q1) to high-density quark phase 
(hereafter Q2) takes place. We recall that the last phase transition 
within the quark phase could be the transition between the 2SC to CFL 
phases, but other options are not excluded. Finally, $s_1$ and $s_2$ 
are squared sound speeds in phases Q1 and Q2, which we quote below 
in units of the speed of light.  

\begin{figure}[tb]
\centering
\includegraphics[width = 0.45\textwidth]{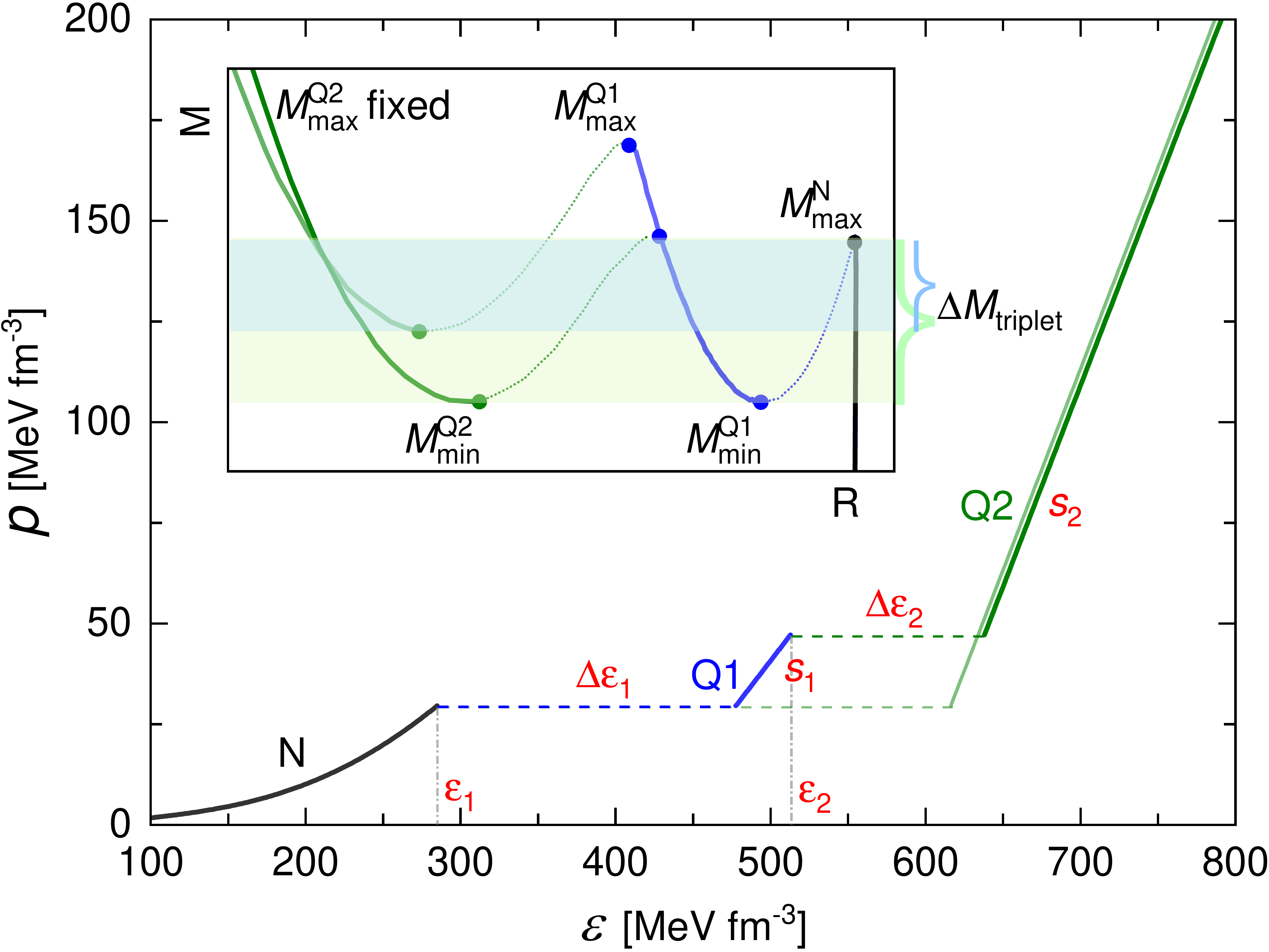}
\caption{ Illustrative EoS and \MR relation for hybrid models. 
Schematic plot showing the parameterizations of the EoS with 
single (light green) and double (dark green) phase transitions 
that could predict the same maximum mass in the quark branch. 
The emergence of triplet configurations that are characterized 
by the minimal or maximum masses in each branch is shown in the 
inset, with one example that obeys conditions~\eqref{eq:cond1} 
and one that does not. 
}
\label{fig:TempoH}
\end{figure}

Figure~\ref{fig:TempoH} provides a schematic picture of the 
double phase transition according to equation~\eqref{eq:EoS}. The six 
independent parameters which enter Equation~\eqref{eq:EoS} are
\bea\label{eq:parameters}
\ep_1, \quad \Delta {\ep_1}, \quad
\ep_2, \quad \Delta {\ep_2}, \quad
s_1, \quad s_2\ .
\eea
For a large enough jump in energy density, $\Delta\ep_1$, when 
the central pressure of a star rises above $p_1$ and Q1 quark 
matter appears in the core, the star becomes unstable.
\footnote{Here and below we use the standard stability criterion, 
which implies that CSs are unstable on the descending branch of 
\MR diagram and stable on the ascending one. This is not always
the case for some boundary conditions on the interface between 
quark and nuclear matter~\citep[see, e.g.,][]{Pereira2018,Goncalves2022,Rau2022}.}
However, it is also possible to regain stability at higher central pressures: 
for a certain range of values of $\ep_1$, $\Delta\ep_1$ and $s_1$, 
there can be a {\it second stable branch} or ``third family'' of CSs. 
In this case, twin configurations may appear, i.e., two stable CSs 
may have the same mass but different radii and, consequently, 
deformabilities. If a second phase transition in the quark phase 
takes place, then a {\it third stable branch} (or ``fourth family'') 
of CSs containing Q2 quark matter in the core can arise. In analogy to 
the above, for suitably chosen parameters, {\it triplet configurations} 
can arise~\citep{Alford2017,Lijj2020a}, in which case there are three 
stable CSs all having the same masses but different radii as well as 
deformabilities.

For the convenience of subsequent discussion, let us define the 
maximal masses of the purely hadronic star (``N star''), the star 
with a Q1 core (``Q1 star''), and the star with a Q2 inner core 
and Q1 outer core (``Q2 star''), as 
\[
M^{\rm{N}}_{\rm{max}},\quad
M^{\rm{Q1}}_{\rm{max}},\quad
M^{\rm{Q2}}_{\rm{max}}\ .
\]
It is also useful to define the minimum values of the masses of the
two stellar branches which consist of stars with only Q1 in the core 
and with Q1 and Q2 in the core, respectively, as 
\[
M^{\rm{Q1}}_{\rm{min}},\quad
M^{\rm{Q2}}_{\rm{min}}\ .
\]
See Figure~\ref{fig:TempoH} for an illustration of these parameters. 
We show also in the inset of this figure schematic \MR relations, which 
illustrate cases where triplets of stars arise, with each new phase of 
matter introducing a new family of CSs. The shaded regions in the 
inset define the range of masses for which triplet configurations arise. 

\begin{figure}[tb]
\centering
\includegraphics[width = 0.45\textwidth]{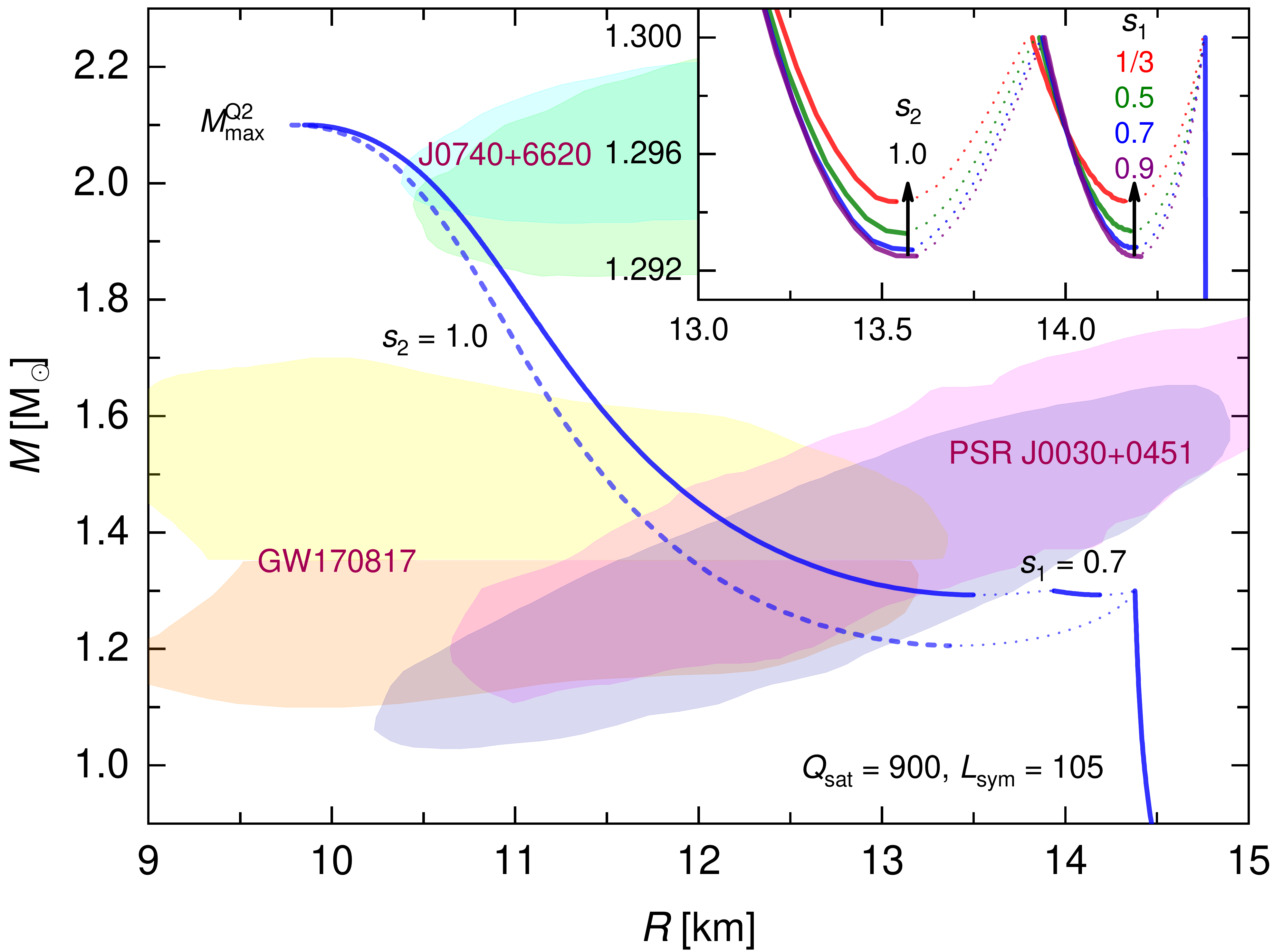}
\caption{
\MR relations showing the stellar sequences have twin or triplet 
configurations in our setup. The inset illustrates the case with a 
fixed $s_2$ value but varying the values of $s_1$. For all curves 
the values of $s_1, s_2$ are as indicated in the plot. The dotted thin 
lines indicate unstable configurations. 
}
\label{fig:SoundE}
\end{figure}

To study the possible role of quark phases in the context of the 
NICER results for the masses, radii, and GW inferences for the TDs, 
we reduce the six-dimensional parameter space~\eqref{eq:parameters} 
as follows:\\
[1ex]
(1) We work in terms of two physical parameters: the maximum mass 
on the nucleonic branch, $M^{\rm{N}}_{\rm{max}}$ (which determines $\ep_1$), 
and the maximum mass on the hybrid branch, $M^{\rm{Q2}}_{\rm{max}}$.
(Note that in the case of a single phase transition we adopt the 
convention of calling the quark phase ``Q2''.) \\
[1ex]
(2) As in~\citet{Lijj2020a}, we impose 
the following conditions (discussed below):
\bea\label{eq:cond1} 
M^{\rm{N}}_{\rm{max}} = M^{\rm{Q1}}_{\rm{max}}, \quad 
M^{\rm{Q1}}_{\rm{min}} = M^{\rm{Q2}}_{\rm{min}}, \quad
M^{\rm{Q2}}_{\rm{max}}\geqslant M^{\rm{Q1}}_{\rm{max}} \ .
\eea
(3) We require the Q2 branch to 
pass through both the PSR J0740+6620 and J0030+0451 ellipses. 
This tightly constrains $s_2$: we find that its value is highly 
correlated with $M^{\rm{Q2}}_{\rm{max}}$.\\
[1ex]
(4) We note that the value of $s_1$ has a negligible influence 
on the \MR curve because the Q1 branch is very short.\\
[1ex]
Using the four constraints/correlations described above, 
we reduce the six-parameter space~\eqref{eq:parameters} to a 
space with two physical parameters, $M^{\rm{N}}_{\rm{max}}$ 
and $M^{\rm{Q2}}_{\rm{max}}$, which specify a nearly 
unique \MR curve.

The conditions~\eqref{eq:cond1} ensure that, if in some range 
of masses twins of the hadronic stars exist due to the phase 
transition to the Q1 phase, the second phase transition leads 
to triplets in the {\it same range} of masses; see the inset 
of Figure~\ref{fig:TempoH} for an illustration. In this inset, 
we show one example that obeys conditions~\eqref{eq:cond1} and 
one that does not, where both models have the same values for 
$M^{\rm{N}}_{\rm{max}}$ and $M^{\rm{Q2}}_{\rm{max}}$.
The inset illustrates the behavior that we typically find: 
if $M^{\rm Q1}_{\rm max}$ rises above $M^{\rm N}_{\rm max}$, 
achieved by increasing $\ep_2$ (namely the width of the Q1 phase)
then at fixed $M^{\rm Q2}_{\rm max}$ we need to reduce the Q1-Q2 
energy-density jump $\Delta\ep_2$ so that $M^{\rm Q2}_{\rm min}$ 
rises above $M^{\rm Q1}_{\rm min}$, which results in a smaller mass 
range for triplets.

In conclusion, the constraint~\eqref{eq:cond1} helps to ensure 
that triplets exist over a reasonable range of masses, 
$\Delta M_{\rm triplet}$, for a hybrid EoS model with given values 
of $M^{\rm{N}}_{\rm{max}}$ and $M^{\rm{Q2}}_{\rm{max}}$. 

Figure~\ref{fig:SoundE} illustrates how transitions to and in 
quark matter allow a nucleonic EoS that is stiff at low densities
to be consistent with astrophysical constraints. In this example, 
the nucleonic EoS has $\Lsym=105$\,MeV (stiff at low density)
and $\Qsat=900$\,MeV (stiff at high density) and we require 
$M^{\rm Q2}_{\rm max}=2.1\,M_\odot$ 
and $M^{\rm N}_{\rm max}=1.3\,M_\odot$, which corresponds to 
a first-order transition at $\rhotran = 1.89\,\rho_{\rm sat}$.
This creates a hybrid branch that is compatible with the GW170817 
and PSR J0740+6620 constraints. As an illustration of the limited 
role of $s_1$, the insert of Figure~\ref{fig:SoundE} shows that 
varying $s_1$ for a fixed $s_2$ changes the mass by $0.001\,M_\odot$.
Note that when we change $s_1$, the parameters $\Delta\ep_1$, $\ep_2$ 
and $\Delta\ep_2$ have to be accordingly adjusted, in order to fulfill 
the first two conditions in Equation~\eqref{eq:cond1}.

\section{Existence of twin or triplet configurations}
\label{sec:Results}
To assess the existence of twin/triplet configurations, we construct 
EoS from the parameter space of our model that allow for mass twins 
or triplets that are consistent with both NICER and GW measurements. 
Specifically, they yield radii that are just above the 90\% CI 
lower limit for PSR J0740+6620 from NICER, and just below the 
90\% CI upper limit on the radius of a $1.36\,M_\odot$ star.
\footnote{Note that $1.36\,M_\odot$ is the mass value inferred for 
an equal-mass binary in the GW170817 event.}
In Section~\ref{sec:light-twins} we will study the stiffer ones that 
require a transition to quark matter if they are to obey the GW170817 
constraint, and in Sections~\ref{sec:heavy-twins} 
and~\ref{sec:massive-twins} we will move on the softer EoS that do 
not require such a transition.

\begin{figure}[tb]
\centering
\includegraphics[width = 0.45\textwidth]{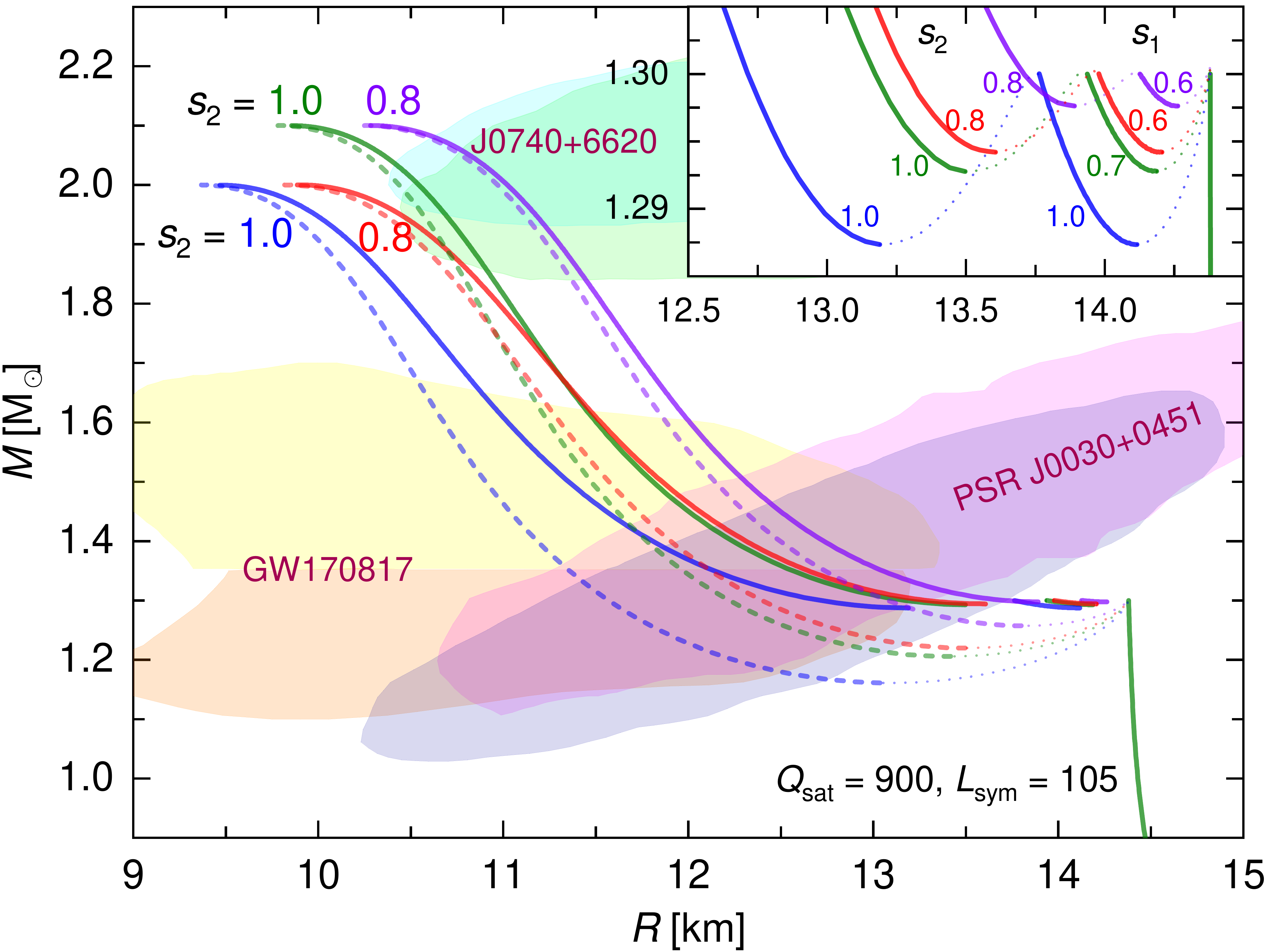}
\caption{
Illustrative \MR relations for hybrid EoS models with stiff nucleonic 
EoS with single (dashed lines) or double (solid lines) phase translations 
at subcanonical masses of star sequences. We use nucleonic EoS 
(with $Q_{\rm sat} = 900, L_{\rm sym} = 105$\,MeV) and
fix $M^{\rm N}_{\rm max} = 1.3\,M_{\odot}$ 
and $M^{\rm Q2}_{\rm max}= 2.0, 2.1\,M_{\odot}$. 
The emergence of subcanonical-mass triplet configurations 
is shown in the inset. For all the hybrid branches, the values 
of $s_1, s_2$ are as indicated in the plot.
}
\label{fig:MR_L}
\end{figure}

\subsection{Hybrid equation of state featuring sub-canonical-mass twins or triplets}
\label{sec:light-twins}
Nucleonic EoS models that are stiff at low and high density 
(with large $L_{\rm sym}$, in the 100\,MeV range, as suggested 
by~\citet{Reed2021}, and positive values of $Q_{\rm sat}$) predict
a large radius for a canonical-mass CS that is in tension with the 
GW170817 inference; see Figure~\ref{fig:TempoN}. We can resolve this 
by positing a first-order phase transition at low density (i.e., in the 
hadronic branch ends below a mass of $1.4\,M_\odot$) leading to a 
hybrid branch of more compact stars. It should be mentioned that the 
same effect can be achieved in the EoS models where heavy baryons like
$\Delta$ resonances appear early. This leads to softening of the EoS 
at intermediate densities, reducing the radius and  TD of the star 
\citep[see, e.g.,][and references therein]{Lijj2020a,Sedrakian2021}.

We have shown previously~\citep{Lijj2021} that twin configurations 
can be expected in the mass interval of $1.2$-$1.4\,M_{\odot}$. 
Having this in mind, we fix $M^{\rm N}_{\rm max} = 1.3\,M_{\odot}$, 
which corresponds to $\rhotran\thickapprox 2\,\rho_\text{sat}$,
and study the general characteristics of the resulting subcanonical-mass 
twins or triplets.

\begin{figure}[tb]
\centering
\includegraphics[width = 0.45\textwidth]{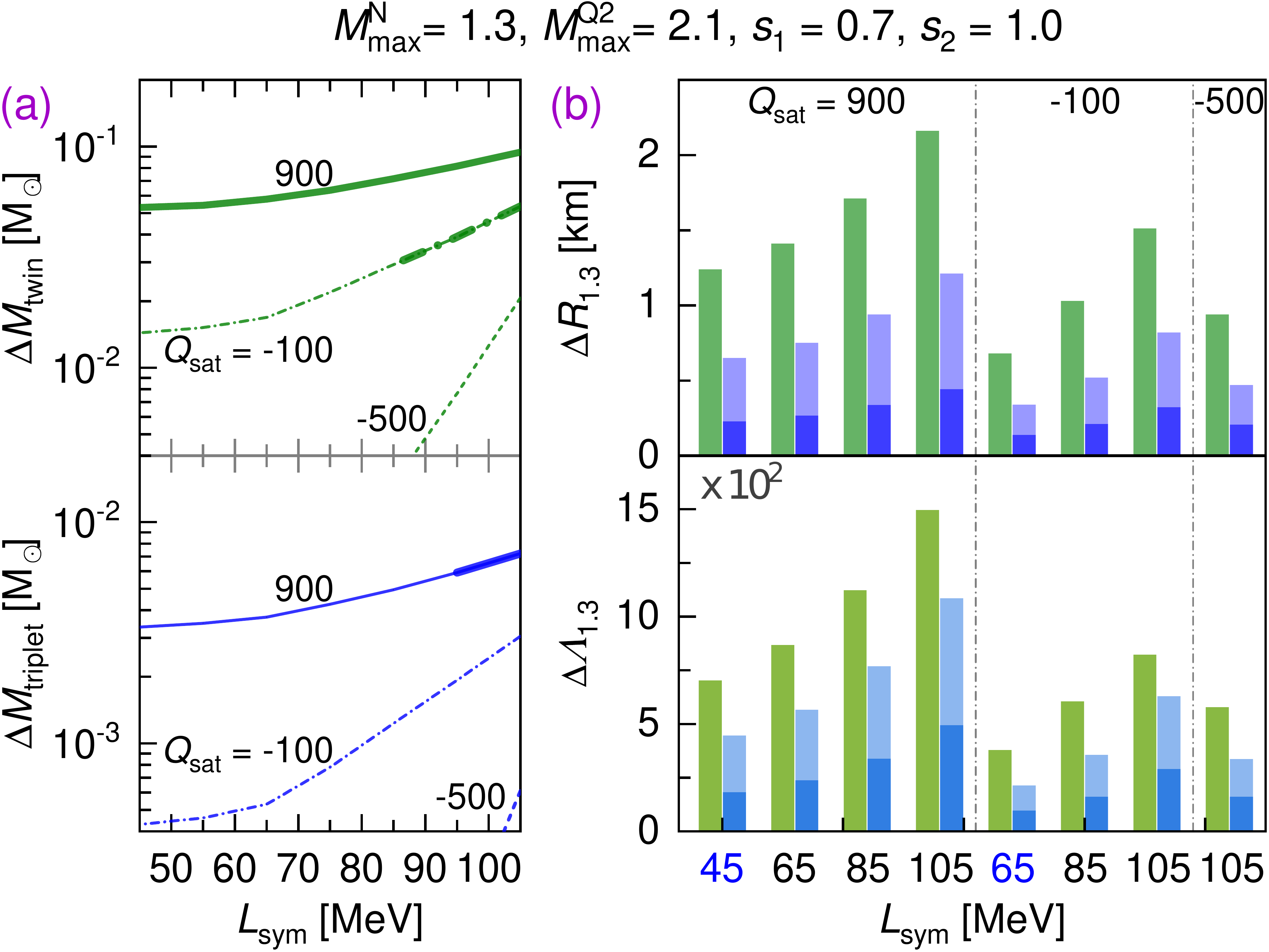}
\caption{
Ranges of parameters characterizing twin and triplet configurations 
for hybrid EoS models.
Panel (a) shows the mass ranges of twins $\Delta M_{\rm twin}$ and 
triplet $\Delta M_{\rm triplet}$;
panel (b) shows the difference in the radius $\Delta R_{1.3}$, and 
TD $\Delta\Lambda_{1.3}$ for twins (in green) and triplets (in blue) 
with mass $M = 1.3\,M_\odot$, as functions of the parameters of the 
nucleonic EoS. In panel (a) those models with 
$\Delta R_{1.3} \geq 1.0$\,km for Q2-N pairs of stars are marked by 
bold lines. In panel (b), for triplets, the column height shows the 
value for Q2-N pair of stars, while the values for Q2-Q1 and Q1-N 
pairs can be read off by the lighter and darker colored column segments, 
respectively. The colored numbers on the x-axis mark those nucleonic 
models that meet GW170817's inference.}
\label{fig:DMR_L}
\end{figure}

Figure~\ref{fig:MR_L} shows exploratory examples based on the
stiffest nucleonic EoS in our collection with 
$Q_{\rm sat} = 900$, $L_{\rm sym}=105$~MeV. We show $M$-$R$ curves 
with two values for the maximum mass on the Q2 branch, 
$M^{\rm Q2}_{\rm max}=2.0$ and $2.1\,M_\odot$, and explore several 
values of $s_1$ and $s_2$. It is seen that the NICER result for 
PSR J0740+6620 puts a strong constraint on $M^{\rm Q2}_{\rm max}$ 
and $s_2$. For $M^{\rm Q2}_\text{max}=2.1\,M_\odot$ it requires 
$s_2 \approx 1.0$ and for $M^{\rm Q2}_\text{max}=2.0\,M_\odot$ 
it requires $s_2 \approx 0.7$. The resultant hybrid stars are also 
compatible with the constraints from GW170817 and the NICER results 
for PSR J0030+0451.

Figure~\ref{fig:DMR_L} summarizes the mass ranges of twins, 
$\Delta M_{\rm twin}$, and triplets, $\Delta M_{\rm triplet}$, 
defined as the range between the maximum value of the nucleonic 
star mass, $M^{\rm N}_{\rm max}$, and the common minimum mass of the 
hybrid star branches, $M^{\rm Q1}_{\rm min}=M^{\rm Q2}_{\rm min}$.
It also gives the difference in the radius, $\Delta R_{1.3}$, and 
TD, $\Delta\Lambda_{1.3}$, for stars with mass $M = 1.3\,M_\odot$, 
as functions of the parameters defining the hybrid EoS model. 

From Figures~\ref{fig:MR_L} and~\ref{fig:DMR_L}, we observe the 
systematic features that were established in the case of only 
twin stars in~\citet{Lijj2021}.

First, increasing $M^{\rm Q2}_{\rm max}$ at fixed $s_2$ or decreasing 
$s_2$ at fixed $M^{\rm Q2}_{\rm max}$ tends to reduce the mass range 
of triplets and twins. At the same time, the instability region of 
mass (or radius) between nucleonic and hybrid stars decreases as 
the value of $M^{\rm Q2}_{\rm max}$ increases at fixed $s_2$, or 
as $s_2$ decreases at fixed $M^{\rm Q2}_{\rm max}$. A smaller value 
of $s_2$, in the current setup, allows a smaller $\Delta\ep_2$ and 
for a resulting steep decrease in the mass range for twins and/or 
triplets. For our parameter choice, in order for the \MR curves to 
pass through the NICER 90\% CI region for PSR J0740+6620, it requires 
$s_2$ close to 1.0.

Second, for stiffer nucleonic EoS featuring larger values of 
$L_{\rm sym}$ and/or $Q_{\rm sat}$, the range of masses where 
twin/triplet stars appear is larger. This is because for hybrid EoS 
with fixed $M^{\rm Q2}_{\rm max}$, increased stiffness of the 
nucleonic EoS must be offset by a softer quark phase, e.g., 
larger energy-density jumps $\Delta {\ep_1}$, $\Delta {\ep_2}$, 
or width $\ep_2-\ep_1-\Delta {\ep_1}$. This makes the hybrid branch 
more compact while the stiffer nucleonic EoS makes the hadronic 
branch less compact, i.e., the difference in radius is larger.
The mass ranges of triplet stars $\Delta M_{\rm triplet}$ are 
typically $\sim 10^{-3}\,M_{\odot}$, which is one order smaller 
than the mass range for twin stars, $\Delta M_{\rm twin}$.

Third, the differences in radii and deformabilities for triplet 
stars are typically 40-50\% smaller than for twin stars. The values 
of $\Delta R_{1.3}$ for N-Q2 pairs are, at most, about 1.0 km; the 
values for N-Q1 or Q1-Q2 pairs are even smaller. Directly observing 
twins or triplets would therefore require a radius measurement accuracy 
of better than 1\,km. This corresponds to a radius measurement 
accuracy of less than 10\%, which is not yet available, but NICER aims 
to achieve 5\% accuracy in the future.
\footnote{NICER Home: https://heasarc.gsfc.nasa.gov/docs/nicer/}
The difference in the TDs of triplets $\Delta \Lambda_{1.3}$ can 
be several hundred to one thousand, so, given TD measurements with 
error bars in the few hundred range, there is a possibility that 
TD measurements from the inspiral phase of CS mergers could identify 
triplets.

\begin{figure}[tb]
\centering
\includegraphics[width = 0.45\textwidth]{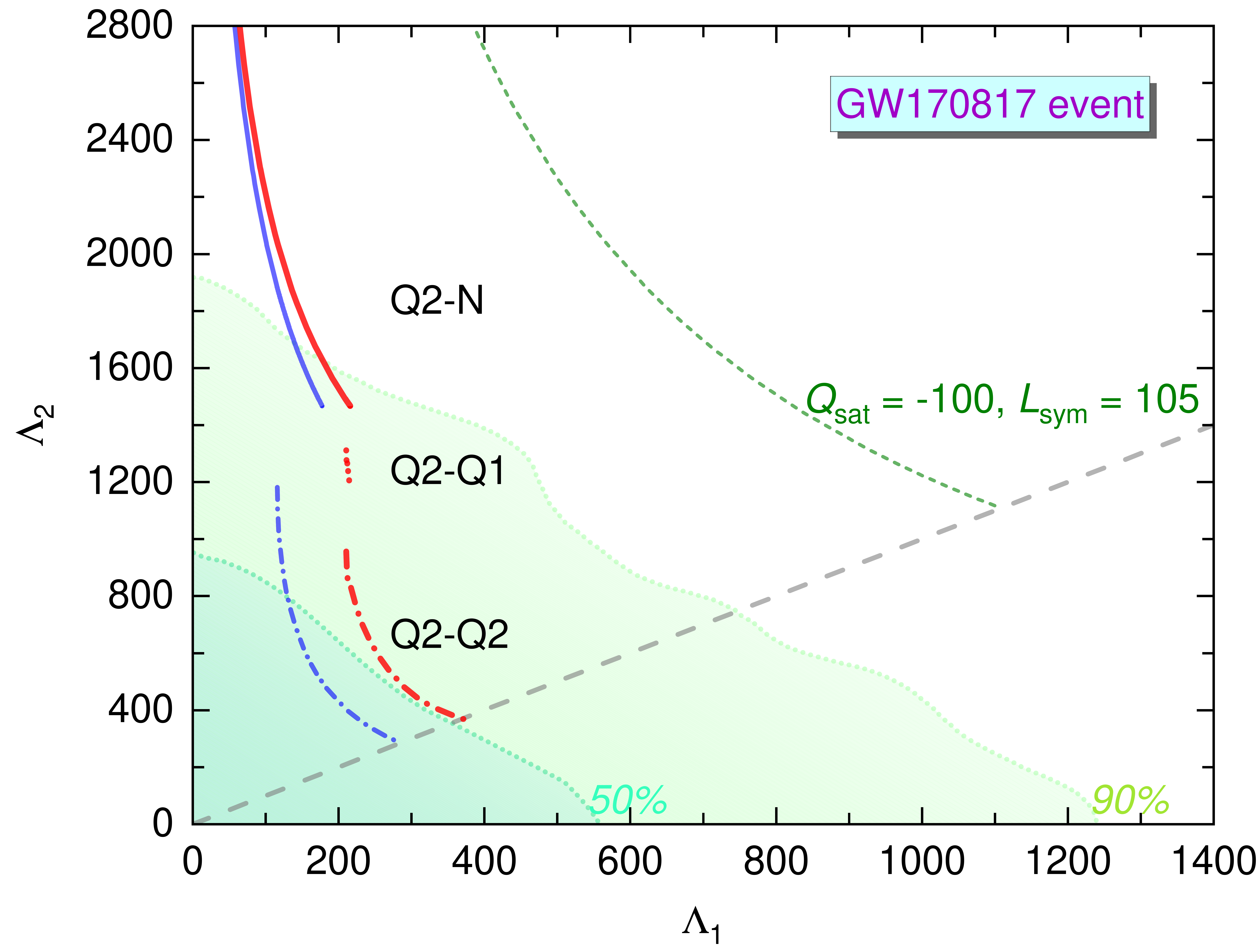}
\caption{
TDs of compact objects with single (in blue) and double (in red) 
phase transitions for a fixed value of binary chirp mass 
$\mathcal{M} = 1.186\,M_\odot$ inferred from the GW170817 event, 
varying the mass ratio. The hybrid models are constructed from 
nucleonic EoS by setting $M^{\rm N}_{\rm max} = 1.3\,M_{\odot}$ 
of the hadronic branch, $M^{\rm Q1}_{\rm max} = 1.3\,M_{\odot}$, 
$s_1 = 0.7$ of the Q1 branch, and $M^{\rm Q2}_{\rm max} = 2.1\,M_{\odot}$, 
$s_2 = 1.0$ of the Q2 branch. For models with mass twins, the 
two types of pairs for stars with masses $M_1$ and $M_2$ are Q2-N and 
Q2-Q1, while for models with mass triplets the three types of pairs 
are Q2-N, Q2-Q1, and Q2-Q2. The shaded regions correspond to the 
50\% and 90\% CIs taken from the analysis of 
GW170817 within the PhenomPNRT model~\citep{LIGO_Virgo2018}.
}
\label{fig:LL_L}
\end{figure} 

We now compare our theoretical TDs for hybrid star models with the 
observational constraints for this quantity obtained from the analysis 
of the GW170817 event~\citep{LIGO_Virgo2018}. We take the chirp mass as 
$\mathcal{M} = 1.186\,M_\odot$ inferred from this merger and carry 
out the comparison only using the analysis which assumes the (more plausible) 
low-spin case~\citep{LIGO_Virgo2018}. For this binary, the component 
masses are found to be in the range $1.16$-$1.60\,{M}_{\odot}$ at 90\% CI. 
In the scenarios that we are exploring in this section, this implies 
that at least one of the components should be a hybrid star.

Figure~\ref{fig:LL_L} displays the TDs $\Lambda_1$ and $\Lambda_2$ of the 
stars involved in the binary with masses $M_1$ (the primary, which is 
defined as the heavier of the pair) and $M_2$ (secondary) in the cases of 
single and double phase transitions. We show curves for two models, 
one with a single Q2 phase and one with both Q1 and Q2 phases.
The models are constructed from soft-stiff nucleonic EoS 
($Q_{\rm sat} = -100$ and $L_{\rm sym} = 105$\,MeV) by setting 
$M^{\rm N}_{\rm max} = 1.3\,M_{\odot}$ of the hadronic branch, 
and combined quark EoS featuring
$M^{\rm Q1}_{\rm max} = 1.3\,M_{\odot}$, $s_1 = 0.7$ of the Q1 branch, 
and $M^{\rm Q2}_{\rm max} = 2.1\,M_{\odot}$, $s_2 = 1.0$ of the Q2 branch.
The diagonal line corresponds to the case of an equal-mass binary with 
$M_{1,2} = 1.362\,M_\odot$. The shaded areas correspond to the 90\% and 
50\% CIs, which are inferred from the analysis of the GW170817 event 
using the PhenomPNRT waveform model~\citep{LIGO_Virgo2019}. 

The hypothesis of a stiff nucleonic EoS with a low-density transition 
to quark matter is compatible with limits on the TD from GW170817.
In Figure~\ref{fig:LL_L} we show how this is possible: for the EoS 
models plotted there, both stars in the GW170817 merger could be hybrid, 
either Q2-Q1 or Q2-Q2, or one could be a heavy hybrid star and the other 
could be a nucleonic star near the top of the nucleonic branch.

\subsection{Hybrid equation of state featuring high-mass twins or triplets} 
\label{sec:heavy-twins}
\begin{figure}[tb]
\centering
\includegraphics[width = 0.45\textwidth]{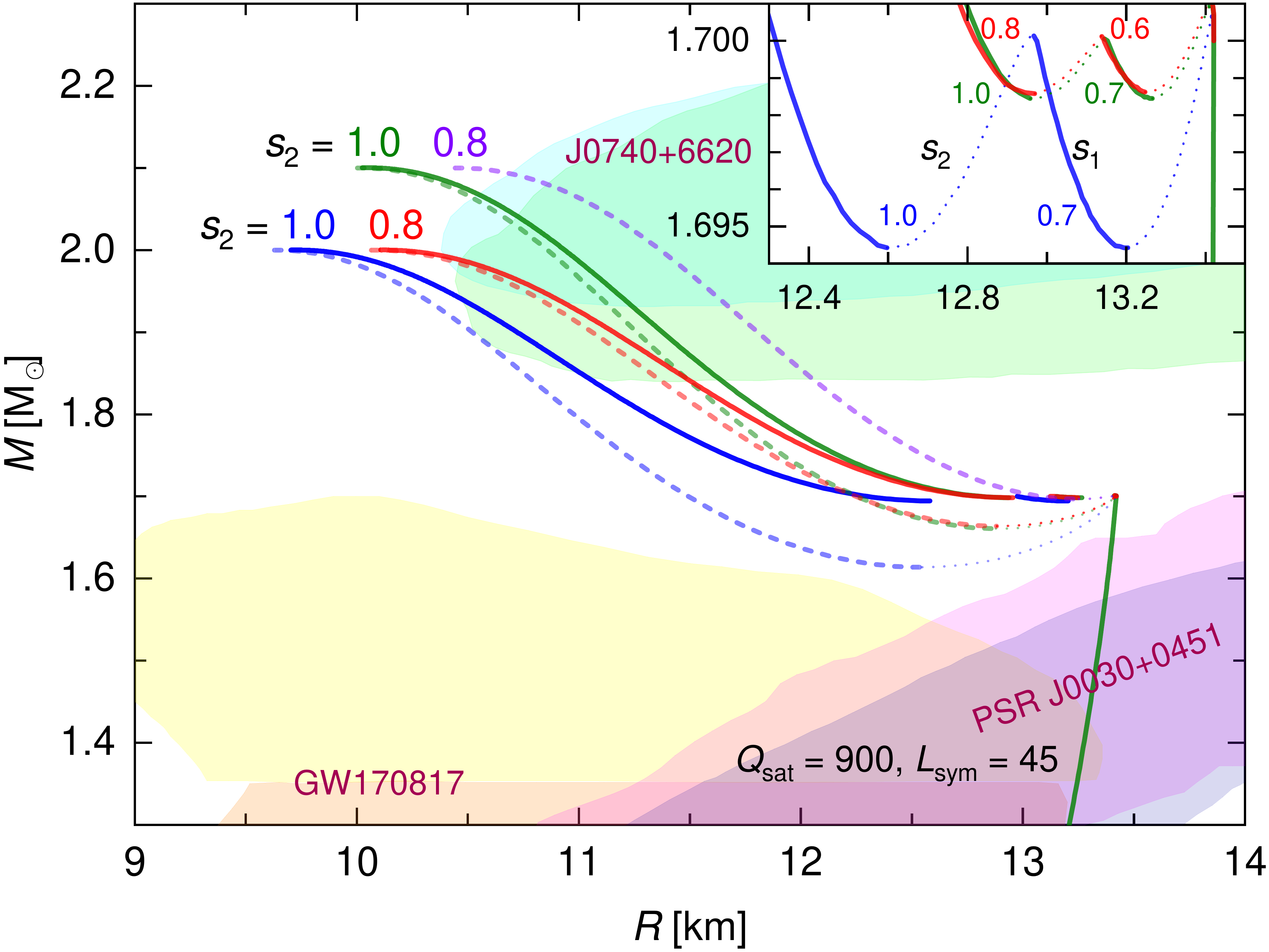}
\caption{
Same as in Figure~\ref{fig:MR_L}, but for hybrid EoS models 
with an intermediate-soft nucleonic EoS that allows single (dashed lines) 
or double (solid lines) phase transitions to appear in high-mass stars.
The results are constructed from a nucleonic EoS 
($Q_{\rm sat} = 900, L_{\rm sym} = 45$\,MeV) by fixing 
$M^{\rm N}_{\rm max}/M_{\odot} = 1.7$ of the nucleonic branch, 
while varying the maximum mass $M^{\rm Q2}_{\rm max}/M_{\odot}= 2.0, 2.1$ 
of the Q2 branch.
}
\label{fig:MR_M}
\end{figure}

Consider next a class of less stiff nucleonic EoS models that 
could (approximately) match GW170817's inference without a phase 
transition to quark matter. Figure~\ref{fig:MR_M} shows examples 
of the nucleonic EoS with $Q_{\rm sat} = 900$, $L_{\rm sym}=45$\,MeV
for a fixed value of  $M^{\rm N}_{\rm max} = 1.70\,M_{\odot}$, which
corresponds to $\rhotran = 2.41\,\rho_{\rm sat}$. The values of the 
remaining parameters --- the sound speed squares, $s_1$ and $s_2$, and 
the maximum mass, $M^{\rm Q2}_{\rm max}$ --- are chosen such that 
the Q2 branch is located close to the lower bound of the NICER's 
90\% CI for the radius of PSR J0740+6620. Within this setup, we find 
the acceptable models are those which have the values of the two 
parameters $M^{\rm Q2}_{\rm max}/M_{\odot}$ and $s_2$ defined as 
pairs, (2.1, 1.0) or (2.0, 0.8), which cover the approximate mass 
ranges for twins and triplets.

\begin{figure}[tb]
\centering
\includegraphics[width = 0.45\textwidth]{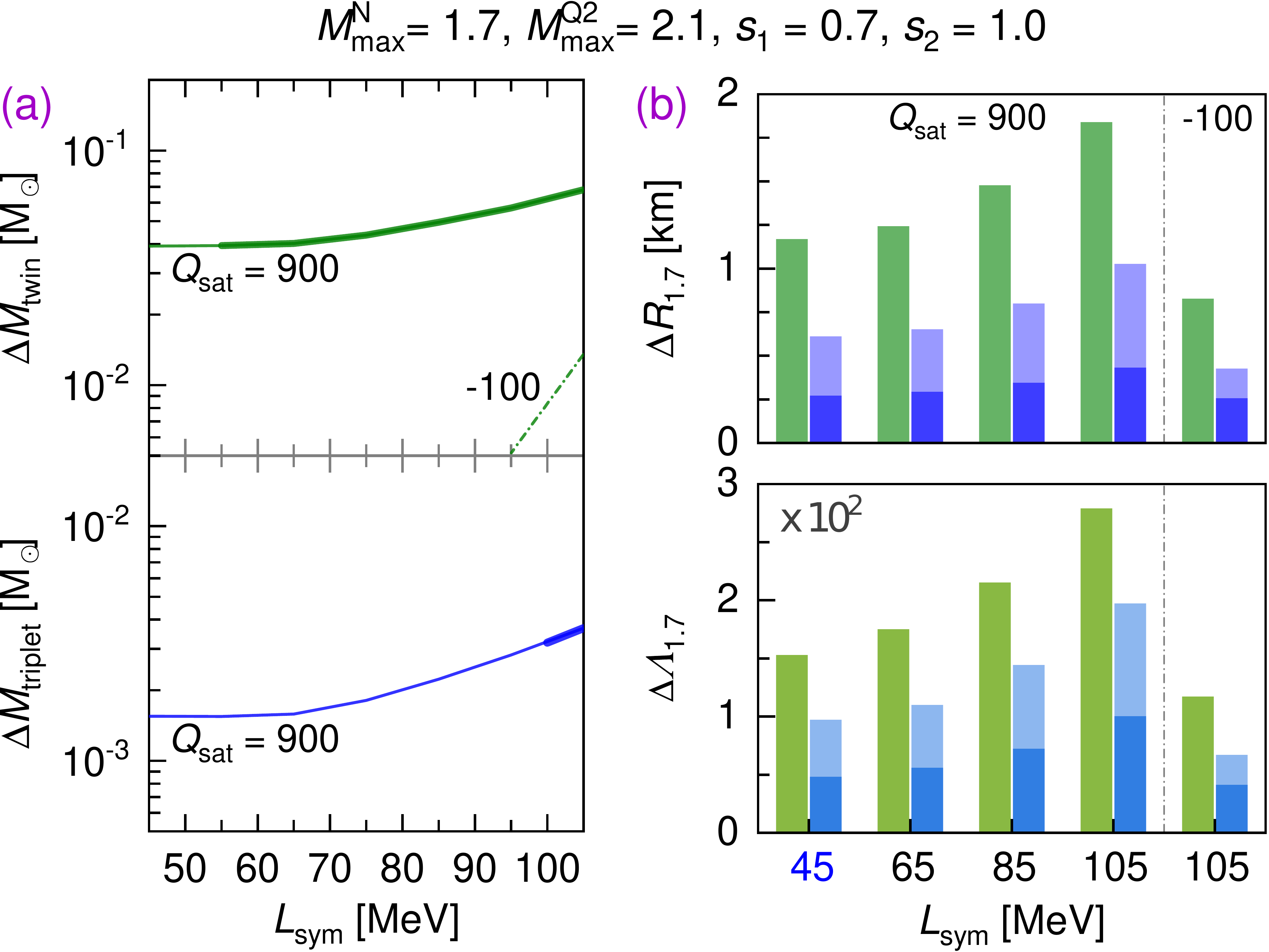}
\caption{
Same as in Figure~\ref{fig:DMR_L}, but for hybrid EoS models 
with $M^{\rm N}_{\rm max}/M_{\odot} = 1.7$. In panel (a) those 
models with $\Delta R_{1.7} \geq 1.0$\,km for Q2-N pairs of stars 
are marked by bold lines. Notice that there are almost no 
twin/triplet configurations for a nucleonic EoS models with 
$Q_{\rm sat} = -500$ and $-100$\,MeV.
}
\label{fig:DMR_M}
\end{figure}

In Figure~\ref{fig:DMR_M} we show the ranges of parameters 
characterizing twin and triplet configurations by varying 
continuously $L_{\rm sym}$ for some fixed values of $Q_{\rm sat}$. 
Note that this figure includes for completeness also the results 
for stiff nucleonic EoS that predict a \MR range outside of the 
GW170817 ellipses.

The general features found for the case of hybrid models featuring 
subcanonical-mass twin/triplet configurations above are replicated 
within this class as well. However, we find (almost) no twin/triplet 
solutions for nucleonic models with $Q_{\rm sat} = -500$, and -100\,MeV; 
see Figure~\ref{fig:DMR_M}\,(a). This implies that the appearance of 
intermediate-mass twin/triplet configurations requires a nucleonic EoS 
which is stiff in the entire relevant range, so that the radius of 
a $1.7\,M_{\odot}$ star $R_{1.7} \gtrsim 13.5$\,km. However, such nucleonic 
EoS all predict $R_{1.4} \gtrsim 13.2$\,km which is inconsistent with 
GW170817's constraint. As a result,  the few valid hybrid EoS models 
in our collection are those constructed from nucleonic EoS with 
$Q_{\rm sat} \sim 900$ and $L_{\rm sym} \sim 45$\,MeV. 
Although we find twin/triplet configurations, as seen in 
Figure~\ref{fig:DMR_M}\,(b), the  differences in their radii 
($\Delta R \lesssim 1$~km) and TDs ($\Delta \Lambda \lesssim 100$) 
are beyond current detection capability.

We now turn to the possible constraints placed by the TDs on our 
collection of hybrid EoS models.  To this end, we set the chirp mass 
$\mathcal{M} = 1.44\,M_\odot$ inferred from the GW190425 
event~\citep{GW190425}. For this binary, the 90\% CIs for the component 
masses range from $1.46$ to $1.87\,{M}_{\odot }$ if we restrict a low-spin 
prior~\citep{GW190425}. The GW190425 observational analysis, however, 
does not provide significantly novel information on the stellar matter 
EoS~\citep{GW190425}. For instance, the estimation of the combined 
dimensionless TD $\tilde\Lambda_{1.44} \lesssim 600$, which could 
be converted to $\Lambda_{1.654} \lesssim 600$ for a star with mass 
$M = 1.654\,M_{\odot}$. This upper limit is consistent with the values 
of TDs predicted by our stiffest nucleonic 
EoS ($Q_{\rm sat} = 900, L_{\rm sym} = 105$\,MeV);
see Figure~\ref{fig:TempoN} where the values of TDs are quoted.

\begin{figure}[tb]
\centering
\includegraphics[width = 0.45\textwidth]{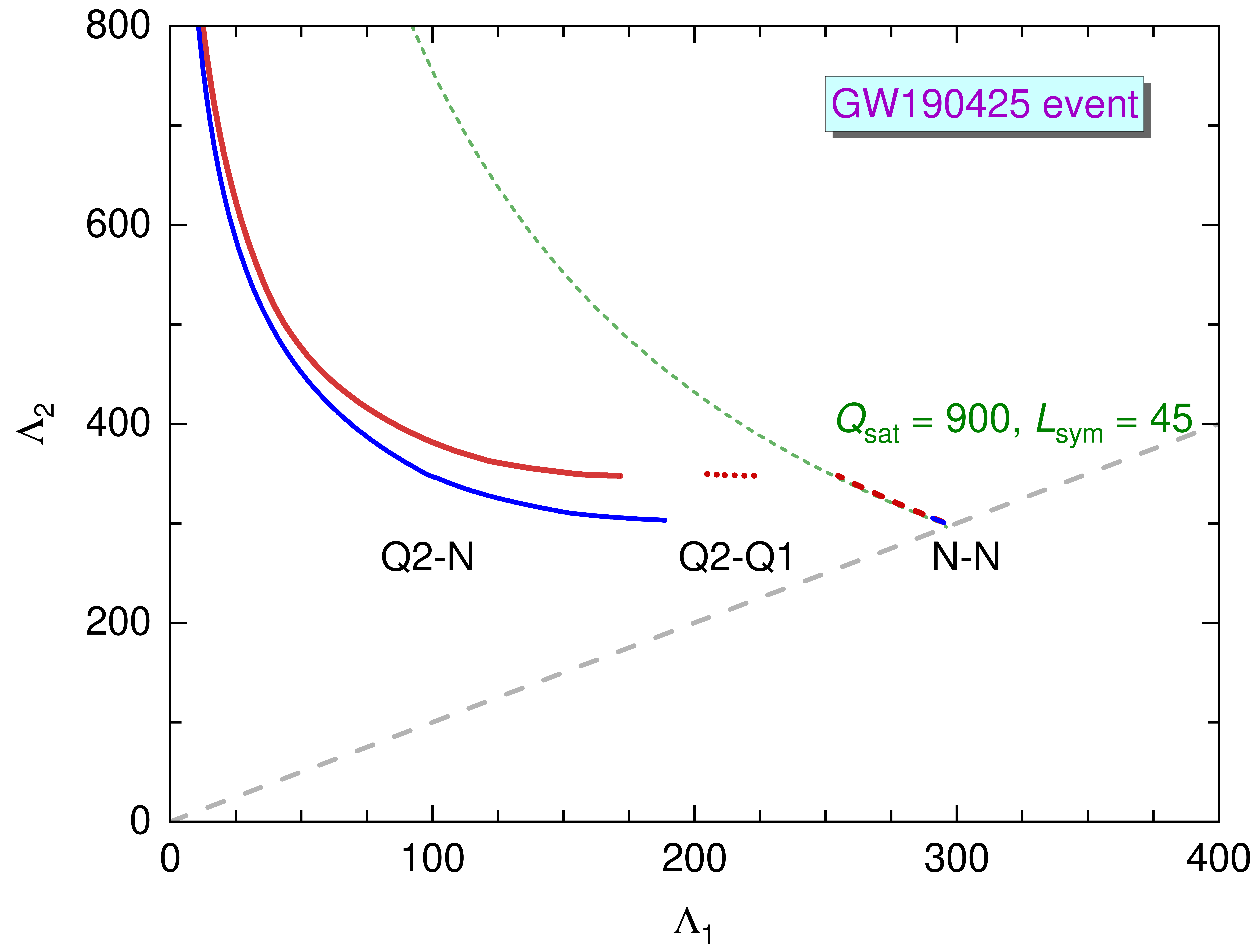}
\caption{
TDs of compact objects with single (in blue) and double phase 
(in red) transitions for a fixed value of binary chirp mass 
$\mathcal{M} = 1.44\,M_\odot$ inferred from the GW190425 
event~\citep{GW190425}. The hybrid EoS models are constructed 
from a nucleonic model with $M^{\rm N}_{\rm max} = 1.7\,M_{\odot}$ 
of the hadronic branch, $M^{\rm Q1}_{\rm max} = 1.7\,M_{\odot}$, 
$s_1 = 0.7$ of the Q1 branch, and 
$M^{\rm Q2}_{\rm max} = 2.1\,M_{\odot}$, $s_2 = 1.0$ 
of the Q2 branch. For models with mass twins, the two types of pairs 
for stars with masses $M_1$ and $M_2$ are Q2-N and N-N, while for models 
with mass triplet the three types of pairs are Q2-N, Q2-Q1, and N-N.
}
\label{fig:LL_M}
\end{figure} 

Figure~\ref{fig:LL_M} displays the TDs $\Lambda_1$ and $\Lambda_2$ 
of the stars involved in a binary with masses $M_1$ and $M_2$ 
in the cases of single and double phase transitions for a selection 
of hybrid EoS. The models are constructed from stiff-soft EoS with 
$Q_{\rm sat} = 900$ and $L_{\rm sym} = 45$\,MeV by setting 
$M^{\rm N}_{\rm max} = 1.7\,M_{\odot}$,
$M^{\rm Q1}_{\rm max} = 1.7\,M_{\odot}$, $s_1 = 0.7$ and 
$M^{\rm Q2}_{\rm max} = 2.1\,M_{\odot}$, $s_2 = 1.0$.
Let us recall that in the case of a single phase transition 
it is assumed that the transition takes place directly from the 
hadronic phase to the Q2 quark phase. The diagonal line corresponds 
to the case of an equal-mass binary with $M_{1,2} = 1.654\,M_\odot$.
In this case the maximum mass of the nucleonic branch of the sequence 
$M^{\rm N}_{\rm max} = 1.70\,M_{\odot}$ is slightly higher than the 
value of the equal-mass case $1.654\,M_\odot$~\citep{GW190425}. The possible 
types of pairs differ from previous results in Figure~\ref{fig:LL_L}. 
In general, two types of pairs of CSs could be involved in such a merger 
event, namely Q2-N and N-N for EoS models with a single phase transition 
featuring mass twins; and three types of pairs of CSs, namely Q2-N, Q2-Q1, 
and N-N for EoS models with two sequential phase transitions featuring 
triplets. As far as observations are concerned, we note that the 
difference in the TDs of triplets $\Delta \Lambda_{1.7}$ for N-Q2 pairs is, 
at most, $\sim$~100. This makes it challenging to distinguish between 
nucleonic and hybrid stars by analyzing the TDs.

\subsection{Hybrid equation of state featuring massive twins or triplets} 
\label{sec:massive-twins}
\begin{figure}[tb]
\centering
\includegraphics[width = 0.45\textwidth]{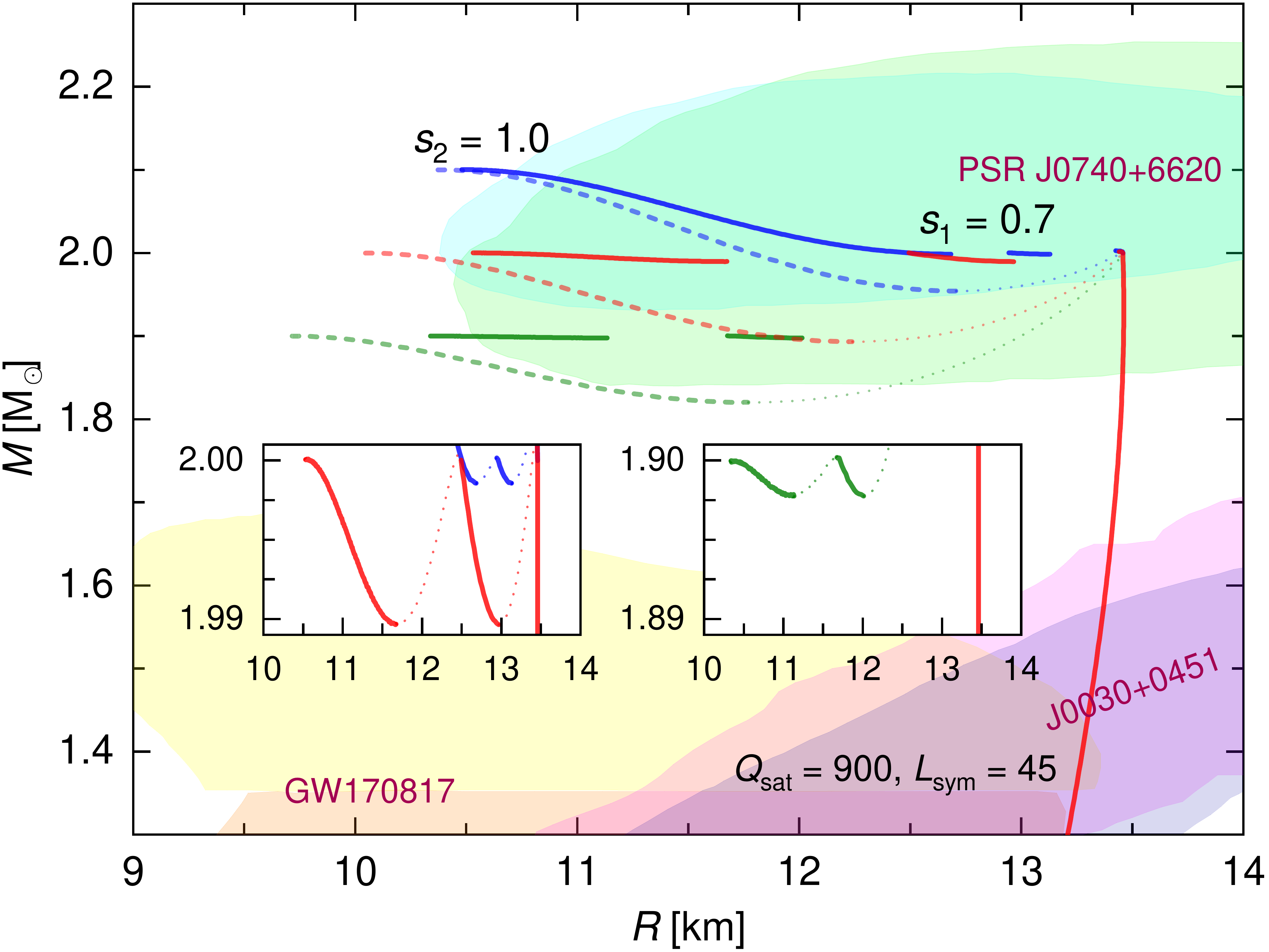}
\caption{
Same as in Figure~\ref{fig:MR_L}, but for hybrid EoS models with 
an intermediate-soft nucleonic EoS that allows single (dashed lines) 
or double (solid lines) phase translations to appear in high-mass stars.
The results are constructed from a nucleonic EoS 
($Q_{\rm sat} = 900, L_{\rm sym} = 45$\,MeV) by fixing 
$M^{\rm N}_{\rm max}/M_{\odot} = 2.0$ of the nucleonic branch, while 
varying the maximum mass $M^{\rm Q2}_{\rm max}/M_{\odot}= 1.9$, 
2.0, and 2.1 of the Q2 branch.
}
\label{fig:MR_H}
\end{figure}

Finally, we consider an extreme class of models where the phase 
transitions occur at high densities $\rhotran \sim 3.0\,\rho_{\rm sat}$, 
i.e., a strong phase transition takes place in CSs of the nucleonic 
branch with a mass close to the value $\sim 2.0\,M_\odot$ of observed 
massive pulsars. In this case, NICER's 90\% CI for PSR J0740+6620 
does not provide significant constraints on the possibility of phase 
transitions. This is because the maximum mass of the nucleonic sequence 
satisfies the requirements set by the NICER measurement and any hybrid 
branches of CSs are allowed to exist even outside the 90\% CI region. 

\begin{figure}[tb]
\centering
\includegraphics[width = 0.45\textwidth]{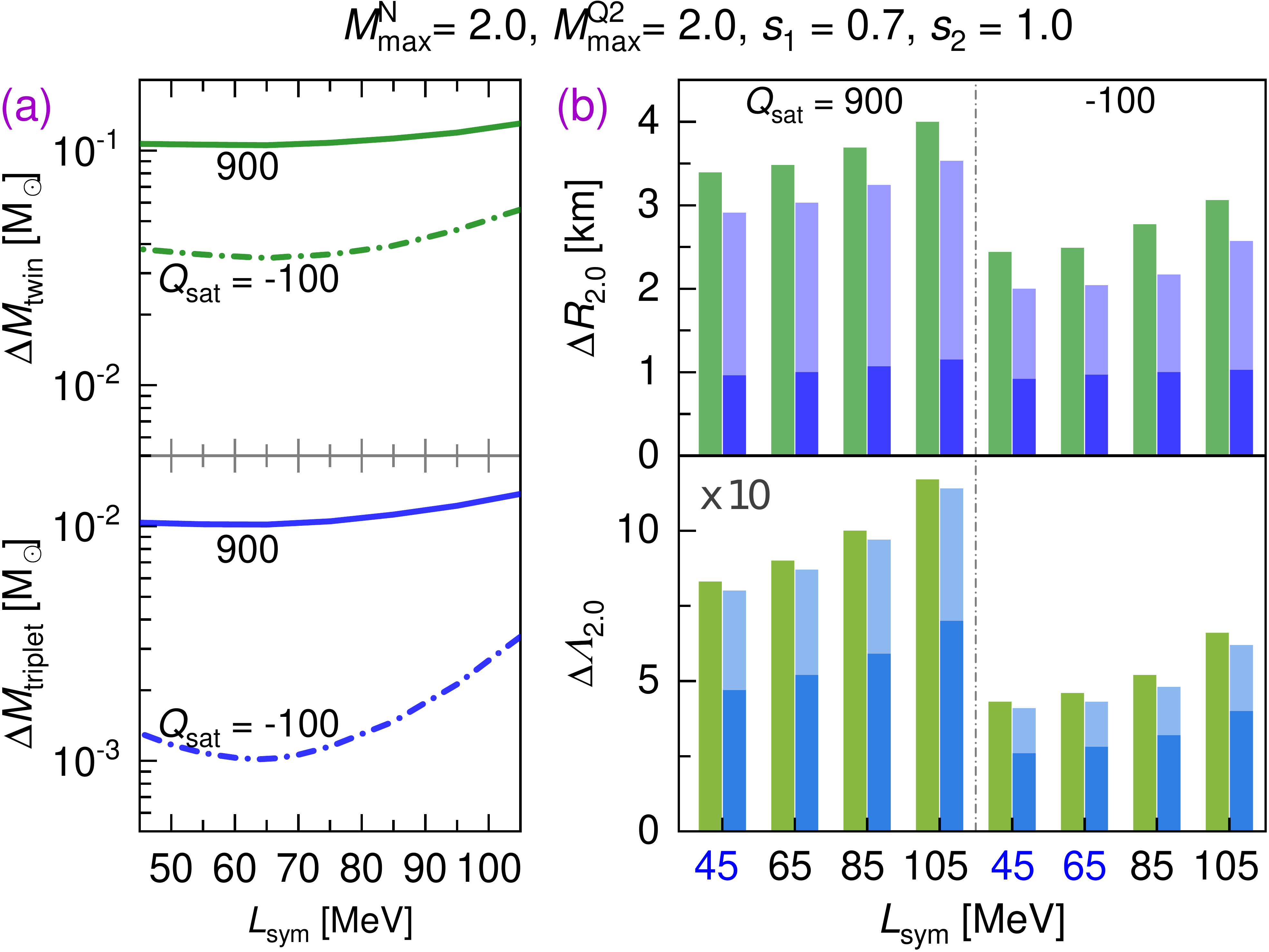}
\caption{
Same as in Figure~\ref{fig:DMR_L}, but for hybrid EoS models with 
$M^{\rm N}_{\rm max}/M_{\odot} = 2.0$. Notice that no twin/triplet 
configurations were found for nucleonic EoS models with 
$Q_{\rm sat} = -500$\,MeV. 
}
\label{fig:DMR_H}
\end{figure}

Figure~\ref{fig:MR_H} shows  model EoS for fixed
$Q_{\rm sat} = 900, L_{\rm sym}=45$~MeV, 
$M^{\rm N}_{\rm max}/M_{\odot} = 2.0$ (which corresponds to 
$\rhotran = 2.76\,\rho_{\rm sat}$), and $(s_1, s_2) = (0.7, 1.0)$
for values of $M^{\rm Q2}_{\rm max}/M_{\odot} = 1.9$, 2.0, and 2.1. 
For the model with $M^{\rm Q2}_{\rm max}/M_{\odot} = 1.9$, 
i.e., a case of $M^{\rm Q2}_{\rm max} < M^{\rm N}_{\rm max}$,
we drop the first condition in Equation~\eqref{eq:cond1} and use 
instead $M^{\rm{Q2}}_{\rm{max}} = M^{\rm{Q1}}_{\rm{max}}$.
It is seen from the insets of Figure~\ref{fig:MR_H} that the most 
pronounced mass twins and/or mass triplets appear in the sequences 
when $M^{\rm Q2}_{\rm max}/M_{\odot} \simeq 2.0$. Figure~\ref{fig:DMR_H} 
gives the ranges of parameters that characterize twin and 
triplet configurations as a function of the model parameters defining 
the EoS. In this class, again, we find no twin/triplet solutions for 
models with $Q_{\rm sat} = -500$\,MeV. 

From Figures~\ref{fig:MR_H} and~\ref{fig:DMR_H}\,(a) we conclude that 
the range of masses containing twin/triplet configurations (if they are
allowed by the parameters of the nucleonic EoS) is somewhat extended 
compared to the $M^{\rm N}_{\rm max}/M_{\odot} = 1.3$ and 1.7 cases 
discussed in previous subsections. The values of $\Delta M_{\rm twin}$ 
could be as large as $0.1\,M_{\odot}$, which is in principle within 
the accuracy of mass measurements via the relativistic Shapiro time 
delay. The radius difference between the radii of twins, 
$\Delta R_{\rm twin}$, can reach up to 2-3\,km, which is again within 
the accuracy that is achieved routinely in NICER data analysis. 
However, in the case of  triplet configurations, the values of 
$\Delta M_{\rm triplet}$ are, at most, $0.01\,M_{\odot}$, which is 
beyond the current detection capability. The radius difference of 
triplets, $\Delta R_{\rm triplet}$, can be as large as 1\,km 
for Q1-N or Q1-Q2 pairs, and 2\,km for N-Q2 pairs; see 
Figure~\ref{fig:DMR_H}\,(b). The differences in the TDs of twins 
and triplets $\Delta \Lambda_{2.0}$, as expected, are only several tens; 
see Figure~\ref{fig:DMR_H}\,(b). This implies that TD measurements 
are not useful for distinguishing very massive nucleonic stars from 
their hybrid twins. We recall that we consider only the class of hybrid 
EoS models for which the nucleonic EoS satisfies the GW170817 constraints.

\subsection{Remarks}
Let us close this section with some general remarks. Combining the 
results presented above, it can be concluded that
(i) if the nucleonic EoS is stiff (i.e., there is tension with GW170817's 
inference) and multiple stable branches with twin or triplet configurations 
exist, then the transition from the nucleonic branch likely happens at a 
low density, $\rhotran \lesssim 2.0\,\rho_{\rm sat}$, with a corresponding 
mass $M^{\rm N}_{\rm max} \lesssim 1.3$-$1.4\,M_{\odot}$; and 
(ii) for a soft nucleonic EoS, which is consistent with GW170817's inference, 
the transition density could be higher, $\rhotran \sim 3.0\,\rho_{\rm sat}$,
with a corresponding mass $M^{\rm N}_{\rm max} \sim 2.0\,M_{\odot}$. 

We found narrow ranges of masses and a modest radius difference for 
low-mass and intermediate-mass twins/triplets, which are a consequence 
of the constraint imposed by the large radius of PSR J0740+6620. It limits 
the allowed range of the reduction of the radius of a hybrid star and makes 
it challenging to distinguish between nucleonic and hybrid stars. The situation 
is more optimistic if the nucleonic branch reaches the maximum value 
$\sim 2.0\,M_{\odot}$. Finally, note that in all the above discussed models, 
a larger mass range for twin/triplet configurations exists for models that 
have a large value of sound-speed square $s_2\le 1$.

\section{Conclusions}
\label{sec:Conclusions}
In this paper, we extended our analysis~\citep{Lijj2021} of the two recent 
observational/experimental results --- the inference of the radius of 
PSR J0740+6620~\citep{NICER2021a,NICER2021b} and the neutron skin thickness 
by analysis of the PREX-II experiment~\citep{Reed2021,Reinhard2021} --- 
to a broader class of models. These now include (i) double phase transitions 
in the quark phase and (ii) transition to quark matter at low, intermediate, 
and high densities. In doing so, we continued using a density-functional 
approach to nucleonic matter which is calibrated to nuclear phenomenology, 
and the CSS parameterization to describe high-density quark matter.

We have assessed the existence of twin or triplet configurations for three 
classes of nucleonic stars. For the class where nucleonic EoS models predict 
a large radius for a canonical-mass ($M=1.4\,M_{\odot}$) star, we show that 
the tension between GW170817's inference for the radius and models 
with large values of $L_{\rm sym}$ can be mitigated (if not completely removed) 
by a low-density phase transition to quark matter and the formation of 
low-mass hybrid stars. This leads always  to the appearance of twins and/or 
triplets, in the cases of single and double  phase transitions, respectively. 
For the class of less stiff nucleonic EoS models that are consistent with 
GW170817's inference, we find that for nucleonic branches extending up to 
$\sim 1.7\,M_{\odot}$, twins or triplets only exist in very small mass ranges, 
i.e., $\Delta M_{\rm twin} \lesssim 0.04\,M_\odot$ and 
$\Delta M_{\rm triplet} \lesssim 0.002\,M_\odot$.

If the nucleonic branch extends up to $\sim 2.0\,M_{\odot}$ 
(i.e., the radius of PSR J0740+662 can be attributed to the nucleonic branch), 
twins and triplets can exist in narrow mass ranges
$\Delta M_{\rm twin} \lesssim 0.1 M_\odot$ for twins and 
$\Delta M_{\rm triplet} \lesssim 0.01 M_\odot$ for triplets. As expected, 
the ranges of mass and radius (and TD) for the existence of mass twins and 
mass triplets are larger for the stiff nucleonic EoS model. The largest ranges 
of twins and triplets are generally supported by models with $s_{1,2}\le 
1.0$, which allow for maximum masses $M^{\rm Q2}_{\rm max} \sim 2.1\,M_{\odot}$.
The most pronounced twins and/or triplet configurations are found to appear in 
sequences with a mass of about $2.0\,M_{\odot}$.

We also extended our previous analysis of the TD~\citep{Lijj2020a,Lijj2021} 
to our current models, with a focus on sequences that contain twin and triplet 
configurations. The resulting TD diagram can be used in future analysis of 
binary neutron star merger events in a search for signatures of QCD phase 
transition(s). In particular, we demonstrated that low-mass twins and triplets 
differ quantitatively by their TDs, while massive twins and triplets differ 
quantitatively by their radii; see Figures~\ref{fig:DMR_L} and~\ref{fig:DMR_H}.
This highlights the prospects of confirmation of the existence of twin and 
triplet stars from future measurements similar to those used in our analysis 
(i.e., X-ray measurements of radii and GW measurements of TDs). 
Specifically, we expect a significantly smaller radius (by about 1-2\,km) for 
hybrid stars compared to purely nucleonic stars. In merger events, finding 
stars with similar masses but significantly different values of TD would be 
a clear indication of a phase transition. In this case, the TDs will be drawn, 
respectively, from the disconnected branches in the $\Lambda$-$\Lambda$ diagrams, 
one corresponding to the nucleonic branch and another to the hybrid branch.
There are other observables that we have not discussed that can indicate a 
first-order phase transition. For example, \citet{Bauswein2019} find that 
the dominant postmerger GW frequency for hybrid and nucleonic 
stars differ significantly from each other. 
Furthermore, the gravitational radiation from asymmetric 
supernova explosions~\citep{Fischer2021,Bauswein2022} will carry 
the imprints of separate phase transitions. Also, matter accretion 
onto a CS will lead to its compression, two successive phase 
transitions and two separate explosions with energy release associated 
with them~\citep{Zdunik2008,Abdikamalov2009,Lin2011}.
Finally, we stress once again that the present astrophysical and 
nuclear physics data do not prohibit the existence of mass twin and/or 
mass triplet if strong first-order phase transitions occur 
in dense matter.\\

J.~L. acknowledges the support of the National Natural Science Foundation 
of China (Grant No. 12105232), the Fundamental Research Funds for the Central 
Universities (Grant No. SWU-020021), and by the Venture \& Innovation Support 
Program for Chongqing Overseas Returnees (Grant No. CX2021007).
A.~S. is supported by the Deutsche Forschungsgemeinschaft Grant 
No. SE 1836/5-2 and the Polish NCN Grant
No. 2020/37/B/ST9/01937 at Wroc\l{}aw University.
M.~A. is supported by the U.S. Department of Energy, Office of
Science, Office of Nuclear Physics under Award No. DE-FG02-05ER41375.


\end{document}